\documentclass[letterpaper, 10 pt, conference]{ieeeconf}  

\IEEEoverridecommandlockouts
\usepackage{xcolor}
\usepackage{soul}
\usepackage{cite}
\usepackage{amsmath,amssymb,amsfonts}
\usepackage{graphicx}
\usepackage{textcomp}
\usepackage{xcolor}
\usepackage{booktabs}

\def\BibTeX{{\rm B\kern-.05em{\sc i\kern-.025em b}\kern-.08em
    T\kern-.1667em\lower.7ex\hbox{E}\kern-.125emX}}
\usepackage[colorlinks]{hyperref}
\usepackage{amssymb}
\usepackage{amsmath}
\usepackage[ruled,longend]{algorithm2e}
\usepackage{multirow}
\usepackage{booktabs}
\usepackage{algpseudocode}

\usepackage{tabularray}
\UseTblrLibrary{booktabs}

\graphicspath{figures/}

\begin{document}

\title{Communication-Aware Consistent Edge Selection for Mobile Users and Autonomous Vehicles}

\author{Nazish Tahir$^1$ \and Ramviyas Parasuraman$^{1*}$ \and Haijian Sun$^2$
\thanks{$^1$ School of Computing, University of Georgia, Athens, GA 30602, USA. \newline
$^2$ School of ECE, University of Georgia, Athens, GA 30602, USA. 
\newline
Authors email: {\tt\small \{nazish.tahir,ramviyas,hsun\}@uga.edu}.
$^*$ Corresponding author.}
}

\maketitle

\begin{abstract}
Offloading time-sensitive, computationally intensive tasks—such as advanced learning algorithms for autonomous driving—from vehicles to nearby edge servers, vehicle-to-infrastructure (V2I) systems, or other collaborating vehicles via vehicle-to-vehicle (V2V) communication enhances service efficiency. However, whence traversing the path to the destination, the vehicle's mobility necessitates frequent handovers among the access points (APs) to maintain continuous and uninterrupted wireless connections to maintain the network's Quality of Service (QoS). These frequent handovers subsequently lead to task migrations among the edge servers associated with the respective APs. This paper addresses the joint problem of task migration and access-point handover by proposing a deep reinforcement learning framework based on the Deep Deterministic Policy Gradient (DDPG) algorithm. A joint allocation method of communication and computation of APs is proposed to minimize computational load, service latency, and interruptions with the overarching goal of maximizing QoS. We implement and evaluate our proposed framework on simulated experiments to achieve smooth and seamless task switching among edge servers, ultimately reducing latency. 
\end{abstract}

\IEEEpeerreviewmaketitle

\section{Introduction}
\label{Sec:intro}
The advent of 5G ultra-communication has exponentially expanded the possibilities and services surrounding autonomous driving. Autonomous vehicles must process vast amounts of data in real time while maintaining high mobility. In the context of Industry 4.0, the importance of computation offloading is driven largely by Internet of Things (IoT) applications. To address these demands, autonomous vehicle systems must utilize a combination of local on-board processing and supplementary remote processing power \cite{7981532}.

To enable these remote processing capabilities, the 5G architecture incorporates multi-access edge computing (MEC). This integration allows for advanced remote processing services that augment autonomous driving with functionalities such as automated traffic management. Here, real-time traffic analysis across an entire area aids vehicles in efficiently navigating to their destinations. Within the literature, use cases that merge vehicular networks with MEC are referred to as vehicular edge computing (VEC) \cite{10.1007/s11036-020-01624-1, 8466364}. It offers an appealing alternative to Cloud Computing by facilitating low-latency, computationally efficient operations while concurrently maximizing system performance. 
 \begin{figure}[htbp]
    \centering
    \includegraphics[width=0.5\textwidth]{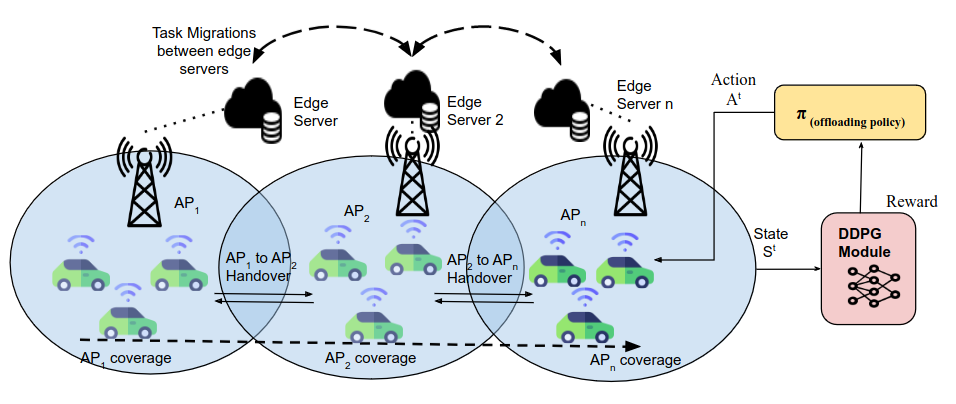} 
    \caption{Communication-aware edge selection in the multi-edge system for a fleet of autonomous vehicles, aided by a DDPG-based learning framework.}
    \label{fig:thematic_diagram}
\end{figure}

The MEC (Multi-access Edge Computing) concept involves replicating traditional cloud data centers' processing and storage capabilities to the network's edge. The base stations in a typical 5G network infrastructure come equipped with edge servers, thus providing additional processing resources as users (or vehicles) need. The computation at the edge ensures there remains ultra-low latency in transferring and processing data at the edge servers due to their close proximity to the users. Leveraging this architecture, vehicular services can operate without resource limitations. 

In the context of offloading time-sensitive tasks to edge devices, the primary challenge in the vehicular environment is the frequent handovers or changes of base stations or access points (APs) due to the vehicle's mobility \cite{pandey2022empirical}. Additionally, selecting and maintaining the optimal communication link for data offloading in fast-moving vehicles presents a significant challenge. When a vehicle exits the coverage area of an AP with co-located edge servers hosting its services, it must re-establish network connections through renewed handshakes and migrate tasks to the appropriate, available, and resourceful computing entity. In a highly dynamic environment, a vehicle's mobility can lead to unnecessary handovers due to slight signal drops, leading to \textit{ping-pong} effect. This process complicates the maintenance of low latency, making it problematic in vehicular edge computing.

Thus, efficient resource management is essential to resolve this issue, which involves strategically selecting access points that offer stable and good communication links to the edge servers and continuously migrating these services as the vehicle moves between access points \cite{parashar2020particle}. This must be done within the constraints of the finite edge resources available at each location. Prioritizing link stability over high link quality ensures less frequent handovers among access points, thereby maintaining service continuity and reducing latency \cite{tardioli2017pound}.

In the context of offloading time-sensitive tasks to edge devices \cite{tahir2023percom}, both computing and communication research communities share a common concern: minimizing AP handovers while ensuring that the vehicle selects the optimal link for data offloading. From a communication standpoint, frequent AP handovers may result in mobile agents always choosing the best quality link, but they can introduce latency due to handshaking and AP re-associations. These delays incur additional computational overhead and contribute to network congestion due to the frequent handover requests. Conversely, the edge/cloud computing community also grapples with significant challenges related to task migrations among the edge servers, which can result in uneven workloads on distributed servers and increased network congestion \cite{li2021multi}.

This paper investigates the integration of learning methodologies to perform offloading decisions, keeping optimal link quality and AP handovers to minimize task migrations and latency.  We present the communication-aware consistent edge selection as a Markovian decision process and propose an approach based on deep reinforcement learning (DRL) off-policy deep deterministic policy gradient (DDPG). Our methodology is tailored to handle continuous task scenarios with a discrete action space. At each time step, an action is chosen, and the resulting outcome is utilized to train the DDPG model to maximize rewards efficiently. The effectiveness of the proposed communication-aware AP-edge selection is demonstrated through extensive simulations and experiments. Results showcase improved task execution times, reduced AP handovers, and enhanced resource utilization, validating the approach's efficacy in dynamic and mobile environments. This research contributes to advancing the design of efficient, intelligent vehicular systems by harnessing edge computing and mobility-awareness for optimized task execution.

\section{Related Work}
\label{sec:relatedwork}

Recently, there has been a shift in addressing resource management challenges in vehicular technology by adopting proactive methods that utilize predictive techniques. These approaches use predictive techniques to reserve edge server resources, ensuring continuous service with minimal delay \cite{7399400,parasuraman2018kalman}. They involve preemptive service placement and migration based on learned mobility patterns and anticipated changes in vehicle locations, optimizing resource usage for seamless service delivery.

Predictive signal modeling \cite{zhang2023map2schedule,parasuraman2023rapid} has been of recent interest in communication-aware robotic and vehicular coordination \cite{ghaffarkhah2010channel,tahir2022analog}.
A predictive approach to solve offloading and migration of computing services is proposed in \cite{9149185} by forecasting user locations and times, balancing resource optimization and latency in multi-user scenarios. However, it overlooks service migrations from frequent handovers and the associated latency. Another work \cite{8745530} optimized computation offloading and resource allocation in collaborative cloud and edge environments but neglected network factors crucial for maintaining connections, thus focusing more on computation than comprehensive resource management.


Transitioning from prediction-based approaches to proactive methods, some research has explored the use of AI algorithms to address optimization and resource allocation challenges in vehicular networks.  For example, in \cite{9548783}, a deep reinforcement learning scheme manages resources. Additionally, \cite{s21020372} proposes a double deep Q network for joint optimization in mobile edge computing (MEC), enhancing computing capabilities while minimizing energy, latency, and communication costs. Similarly, \cite{9026875} explores spectrum, computing, and storage resource allocation using reinforcement learning in MEC-based vehicular networks to meet quality-of-service demands efficiently.

Researchers have explored multi-dimensional resource management for UAV-assisted vehicular networks. In \cite{9456882}, a multi-agent deep reinforcement learning framework is proposed to optimize channel allocation and power control in heterogeneous vehicular networks, enhancing convergence and system performance for diverse QoS needs. Similarly, authors in \cite{9254093} used a multi-agent DDPG-based method optimizes MEC server resource allocation for efficient task offloading and QoS requirements. Similarly, authors 

In \cite{8485780}, researchers used a modified Genetic Algorithm for optimizing offloading, path-planning, and AP-selection under time and energy constraints. \cite{Bakhtiarnia2022DynamicSC} addressed dynamic split computing between mobile devices and edge servers, optimizing task offloading based on network conditions and device mobility. Hayat et al. \cite{9363523} studied edge computing in 5G for autonomous drone navigation, comparing image processing offloading modes: onboard, fully offloaded, and partially offloaded.



In the work of Saboia et al.\cite{9837416}, a multi-layer networking solution enhances scalability and bandwidth efficiency for multi-robot systems through bandwidth-aware prioritization. Another work by Ghiasi et al.\cite{9849036} address AP selection in cell-free massive MIMO systems, optimizing association parameters while considering practical constraints like training errors and access to channel state information.

The above literature survey underscores a significant gap in research concerning minimizing handovers and optimizing resource allocation in mobile vehicle contexts. Our framework addresses this by introducing a resource allocation model that optimizes AP selection based on signal quality and load balancing, aiming to reduce frequent AP handovers and minimize service migrations.

Accordingly, the key contributions in this paper include:
\begin{itemize}
\item Establishing a collaborative vehicular network with access points linking edge servers and vehicles for parallel computing of delay-sensitive tasks.
\item Proposing a joint optimization approach using a DDPG-based framework to enhance network stability, AP handover, and load balancing while meeting QoS requirements of uninterrupted services.
\item Establishing optimal vehicle-AP associations through offline training to meet time-sensitive QoS needs.
\item Evaluating the algorithm through simulations, showing superior performance compared to baseline methods.
\end{itemize}

\begin{figure*}[t]
    \centering
    \includegraphics[width=.90\textwidth]{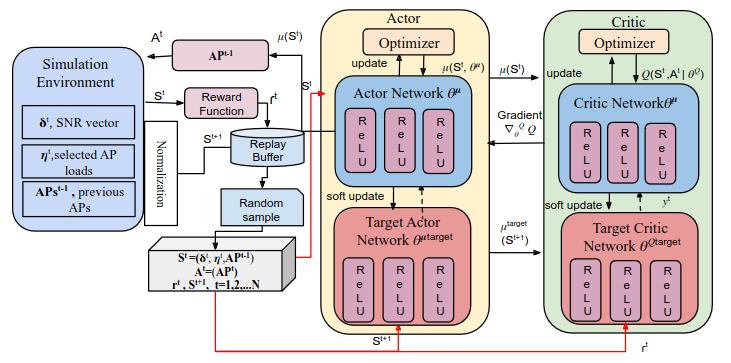}
    \vspace{-2mm}
    \caption{\textcolor{black}{DDPG-based Communication-aware Edge Selection  Framework for a Fleet of Autonomous Vehicles.}}
    \label{fig:mainfig}
    \vspace{-2mm}
\end{figure*} 

\section{Methodology}
\label{sec:method}
\subsection{Overview of Deep Reinforcement Learning}
Deep Reinforcement learning is a type of machine learning paradigm for automated decision-making through interactions with the environment \cite{sutton2018reinforcement}. The vehicular community has extensively utilized Deep Reinforcement Learning due to its efficacy in computing service latency and improving service reliability for
vehicular network \cite{9119487}.

In a typical reinforcement learning problem, the objective of an agent is to effectively determine a strategy (policy) that maximizes the long-term reward, where the learning problem can be modeled as a Markovian Decision Problem (MDP) \cite{ernst2024introduction}. MDP is a tuple $(S, A, \rho, f, \gamma)$, explained below.
\begin{enumerate}
    \item $\mathcal{S}$ represents the state space, i.e., a set of all observable states.
    \item $\mathcal{A}$ represents the action space, i.e., a set of all valid actions. The actions can be discrete or continuous. 
    \item $\pi:S \rightarrow \mathcal{A}$ represents the probability distribution with which the agent samples the next action, $a$ if the state changes from s to the next state $s'$.
    \item $\rho:\mathcal{S}\times \mathcal{A} \rightarrow \mathbb{R}$ is the immediate scalar reward that the agent observes after taking the action a, transitioning the environment from s to s'. 
    \item $f :\mathcal{S}\times \mathcal{A} \times \mathcal{A} \rightarrow [0, 1]$ represents the state transition function. $f (s, a, s') = P(s' |s, a)$ is the probability that state $s$ transits to $s'$ after action $a$ is performed.
    \item $\gamma \in [0, 1]$ is a discount factor that reduces the impact of future rewards on the present \cite{9904958}.
\end{enumerate}

The reward $R_{t}$ of MDP is formulated as 
\begin{equation}
\label{eq:MDP}
    G_t = \sum_{k=0}^{\infty} \gamma^k R_{t+k+1} ,
\end{equation} 
where $R_{t+k+1}$ represents the immediate reward obtained at time step $t+k+1$, and $\gamma$ is the discount factor. 

The value function determines how good it is for the agent to be in a given state, and a state-value function is quantified by the cumulative reward in a particular state following a policy $\pi$.
\begin{equation}
    V^\pi(s) = \mathbb{E}_\pi \left[ G_t \mid S_t = s \right] = \mathbb{E}_\pi \left[ \sum_{k=0}^{\infty} \gamma^k R_{t+k+1} \mid S_t = s \right]
    \label{eq:state-value}
\end{equation}

Similarly, an action-value function (Q-function) determines the cumulative reward of taking an action $a$ in state $s$ following a policy $\pi$ .
\begin{equation}
\label{eq:action-value}
    Q^\pi(s, a) = \sum_{s' \in \mathcal{S}} \mathbb{P}(s' | s, a) \left[ \mathcal{R}(s, a, s') + \gamma V^\pi(s') \right]
\end{equation}
The agent selects an action based on the policy, which could be stochastic, yielding a probability distribution of an action on a given state space $\pi(s, a)$ or deterministic, leading to an action with certainty. The ultimate goal is to find the optimal policy that maximizes the total cumulative reward and, for that, the optimal state-action value function that indicates the maximum possible reward for each action given a specific state given as $\max_a Q^*(s, a) = V^*(s)$. Thus, the optimal state-action value function that the agent aims to find can be defined as: 
\begin{equation}
\label{eq:state-action}
    Q^*(s, a) = \sum_{s' \in S} P(s'|s, a)[R(s, a, s') + \gamma \max_{a'} Q^*(s', a')]
\end{equation}

\paragraph*{Deep Deterministic Policy Gradient (DDPG)}
DDPG \cite{lillicrap2015continuous} is a model-free off-policy actor-critic algorithm designed for continuous environments that employ neural networks to approximate the Q-value for each state and action pair through a critic network (parameterized by $\theta_c$) and an actor-network (parameterized by $\theta_{\pi}$) to estimate optimal actions. This actor-critic architecture is well-suited for continuous action spaces. The critic network is trained similarly to Deep Q-Network, while the actor-network is updated using policy gradient by applying the chain rule.

\subsection{SNR calculation}
Traditional approaches for AP selection and handover are typically based on the measurement of Signal-to-Noise Ratio (SNR). SNR is used to identify candidate APs and ensure the availability of wireless links. An ideal AP selection algorithm should maintain continuous service and balance the network's overall load. Conventional methods select the AP with the strongest SNR without considering the traffic loads among APs, which often results in significantly uneven traffic distribution. Instead of solely selecting an AP with the maximum SNR, our proposed learning model incorporates the traffic load on each access point into the state of the environment during the learning process.

The SNR calculation is performed by initializing the transmission power ($\textit{txPw}$) to 25 mW and the operational bandwidth ($\textit{opBW}$) to $2 \times 10^7$ MHz. The channel parameters are specified in the Table \ref{table:snr_parameters}. 

The noise power ($\text{nPw}$) is calculated using the formula:
\begin{equation}
\label{eq:noise-power}
\textit{nPw} = 10^{\frac{7}{10}} \times 1.3803 \times 10^{-23} \times 290 \times \textit{opBW}
\end{equation}

For each AP, the path loss ($\textit{Ploss}$) is computed based on the distance between the AP and the vehicle. If the distance is less than the breakpoint distance ($\textit{dBP}$), the path loss is determined using the parameters before the breakpoint; otherwise, it is computed using the parameters after the breakpoint. Additionally, random noise is added to account for shadow fading. The total path loss is then adjusted by adding penetration losses and fixed shadow fading. 

Finally, the signal-to-noise ratio (SNR) is calculated as
\begin{equation}
\label{eq:snr}
\textit{snr\_v} = \textit{txPw} - \textit{Ploss} - 30 - 10 \times \log_{10}(\textit{nPw})
\end{equation}

These SNR values are then normalized by scaling the SNR vector between minimum and maximum SNR values.

\begin{table}[h!]
\centering
\caption{Parameters for SNR Calculation}
\begin{tabular}{p{4cm} p{4cm}}
\toprule
\textbf{Parameter} & \textbf{Value} \\
\midrule
Transmission Power (txPw) & 25 mW \\
Operational Bandwidth (opBW) & $2 \times 10^7$ Hz \\
Frequency & 5 GHz \\
Fixed Shadow Fading Temperature & 18 dB \\
Channel Type  & B \\
Channel Line-of-Sight (Liw) & 7 dB \\
Breakpoint Distance (dBP) & 5 m \\
Pre-Breakpoint Slope  & 2 \\
Post-Breakpoint Slope & 3.5 \\
Line-of-Sight Attenuation Before Breakpoint  & 3 dB \\
Line-of-Sight Attenuation After Breakpoint  & 4 dB \\
\bottomrule
\end{tabular}
\label{table:snr_parameters}
\end{table}

\section{Implementation Details}  

\textbf{System Model:} 
Suppose there are $N$ vehicles denoted as V = $\{V_{1}, V_{2}, V_{3}, ....V_{N}\}$, and each vehicle has a computationally intensive task needing to be offloaded. There are multiple distributed Wi-Fi Access Points ($AP$s), and these $AP$s form the edge nodes, which serve as the remote computational resources available to the vehicles at the edge network. While autonomously navigating towards the target, each vehicle $V$ must determine the optimal access point ($AP$) with the most suitable network resources to offload its computationally intensive service. Typically, due to mobility and overlapping coverage of the access point, the vehicles need to make handovers. 

Our proposed RL module uses signal strength measured as SNR, and each access point loads as input to perform intelligent AP selection. The RL module learns, based on the vehicle's location, AP-load information, and network control information (signal-to-noise ratio or SNR), the best $AP$ for the vehicle to ensure reliable data transfer for maintaining the QoS for time-sensitive tasks. In this paper, the RL agent's co-optimization framework can balance communication quality and stability, ensuring the least handovers while maintaining performance guarantees.

Given the constraints of the proposed scenario, we formulate a DRL network resource orchestrator that aims to learn the optimal network resource orchestration over time. The respective orchestrator ensures the vehicle reaches its intended target location by connecting to the right $AP$ with maximum SNR and optimized AP-to-vehicle load while ensuring minimum handovers possible, which would ensure correct and timely delivery of the vehicle's sensor data to the edge for offloading services to satisfy the application QoS. 

 \textbf{\textit{Problem 1 }}--- 
The objective is to maximize the cumulative utility $U_i$ for each vehicle $V_i$ by selecting optimal access points while satisfying SNR thresholds and access point load constraints. The objective function is formulated as follows:
\begin{equation}
\label{eq:obj_func}
\begin{aligned}
\max \quad & \sum_{i=1}^{N} U_i \\
\text{s.t.} \quad & \text{SNR}_{ij}(t) \geq \text{SNR}_{\text{th}}, \quad \forall i, j, t \\
& \text{Load}_{j}(t) \leq \text{MaxLoad}_{j}, \quad \forall j, t ,\\ 
\end{aligned}
\end{equation}
where $U_i$ is the cumulative utility of vehicle $V_i$, $N$ is the total number of vehicles, $\text{SNR}_{ij}(t)$ is the Signal-to-Noise Ratio (SNR) between vehicle $V_i$ and access point $AP_j$ at time $t$, $\text{SNR}_{\text{th}}$ represents the minimum acceptable SNR threshold, $\text{Load}_{j}(t)$ is the load on access point $AP_j$ at time $t$, indicating the number of connected vehicles, and $\text{MaxLoad}_{j}$ is the maximum load that access point $AP_j$ can handle. 
Additionally, $U_i$ can be further defined as presented in Eq.~\eqref{eq:obj_func_defined}
\begin{equation}
\label{eq:obj_func_defined}
\begin{aligned}
U_i = & \sum_{t=1}^{T} \gamma^{t-1} \left( \alpha \cdot u_{\text{SNR}}(i, t) + \beta \cdot u_{\text{Load}}(i, t) \right.\\
& \left. - \gamma \cdot u_{\text{Handover}}(i, t) + \delta \cdot u_{\text{Target}}(i, t) \right) ,
\end{aligned} 
\end{equation}
where (\(\gamma\)) represents the discount factor, with \(0 \leq \gamma \leq 1\), employed to discount future utilities, aligning with the principle that immediate utilities typically hold greater value than those in the distant future. The term \(\gamma^{t-1}\) ensures a weighted consideration of utilities closer to the present time.

The SNR utility (\(u_{\text{SNR}}(i, t)\)) characterizes the utility associated with the Signal-to-Noise Ratio (SNR) for vehicle \(V_i\) at time \(t\). A higher SNR typically indicates superior connection quality, thereby positively contributing to the overall utility. The weight \(\alpha\) underscores the significance of SNR in the aggregate utility.

The load utility (\(u_{\text{Load}}(i, t)\)) gauges the utility contingent upon the load on access points. This utility is influenced by the number of vehicles connected to a given access point, aiming to equitably distribute the load across access points. The weight \(\beta\) denotes the importance assigned to this factor.

The handover utility (\(u_{\text{Handover}}(i, t)\)) penalizes instances of handovers, occurring when a vehicle transitions from one access point to another. Frequent handovers can diminish service quality and amplify operational costs within the network. Thus, the weight \(-\gamma\) (with \(\gamma\) serving as the penalty coefficient) reflects the adverse impact of handovers.

The target utility (\(u_{\text{Target}}(i, t)\)) increases as vehicles reach or progress toward their designated locations. Reaching the destination holds paramount importance in applications such as navigation and logistics. The weight \(\delta\) underscores the significance of this utility in the overall objective.

\textbf{Solution}
The overall framework of the proposed DDPG RL agent for solving problem P is illustrated in Fig.~\ref{fig:mainfig}.  The RL agent employs a Markovian Decision Problem (MDP) for its goal of selecting a policy that maximizes the total reward it receives when interacting with an environment.  
The objective of the RL agent is to tune its policy with the ultimate goal of making the vehicle reach the target by selecting the best AP with respect to SNR and AP load metrics. The policy updates $\pi_\theta$ occur iteratively at each time slot $t$ dynamically adjusting the actor-critic network, based on the collected reward $r^t$ and the observed state $\mathcal{S}^t$ of the vehicle generated by the environment $\psi$ following the execution of the action $\mathcal{A}^t$ on it. 
Below, we detail the state, action, and reward spaces. 

\paragraph*{State Space}
Our state space is defined as the combination of the associated SNR specified in a matrix $\delta^t$ of each vehicle and the available APs and $\eta^t$, AP's Load specifying the computational utilization of the current AP by the numbers of vehicles connected to each AP. We also include the AP selections of the vehicles $AP_{t-1}$ from the previous time step into the state vector.  
We formulate the state space as $\mathcal{S}^t = \{{\eta^t , \delta^t, AP^{t-1} }\}$.

\paragraph*{Action Space}
Action space comprises a vector that defines the AP-vehicle associations at each time step. Therefore, the action of the network orchestration is defined as     $\mathcal{A}^t = \{{AP^{t}}\} $.

\paragraph*{Reward Space}
The reward function is defined as
\begin{equation}
    r^t = f(\mathcal{S}^t, \mathcal{A}^t) = r_{SNR} + r_{AP_{load}} + r_{handover} .
    \label{eqn:reward}
\end{equation}
Here, $r_{SNR}$ represents the positive reward for the least difference between the selected AP's SNR to the greedy-SNR-based AP selection (Eq.~\eqref{eq:reward_snr}),  $r_{AP_{load}}$ is the positive reward for low loads for the selected APs (Eq.~\eqref{eq:r_AP_load}), and $r_{handover}$ an exponentially negative reward for increasing handovers (Eq.~\eqref{eq:r_handover}).
\begin{equation}
\label{eq:reward_snr}
    r_{\text{SNR}} = 2 \left( \frac{{e^{\text{curr\_snr}} - e^0}}{{e^1 - e^0}} \right) - 1
\end{equation}
  \begin{equation}
    \begin{aligned}[b]
\label{eq:r_AP_load}
r_{\text{AP}_{\text{load}}} = 
\begin{cases} 
\text{penalty} + \exp\left(10 \left(\frac{\text{curr\_load}_i - \text{max\_load}}{\text{max\_load}}\right)\right) & \\\text{if } \text{curr\_load}_i < \text{max\_load} \\
-1 & \\ \text{if } \text{curr\_load}_i = \text{max\_load}
\end{cases}
\end{aligned}
  \end{equation}
\begin{equation}
\label{eq:r_handover}
r_{\text{handover}} = - e^{-k \cdot \text{handovers}}
\end{equation}
Lastly, a termination reward is also applied based on the outcomes.
\begin{equation*}
    r^t = r_{termination} = \begin{cases} 
      2500 & \text{if } outcome = SUCCESS \\
      -2000 & \text{if } outcome \neq SUCCESS \\
      0 & \text{otherwise} 
\end{cases}
\end{equation*}

\begin{figure*}[t]
    \centering
    \includegraphics[width=.95\textwidth]{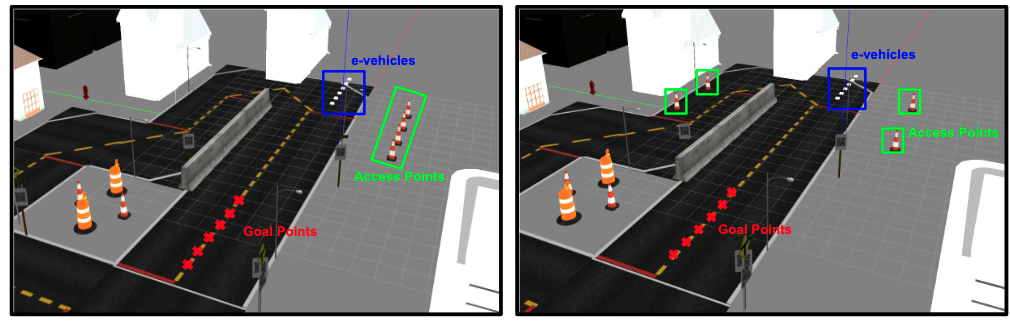}
        \vspace{-2mm}
    \caption{\textcolor{black}{The simulation setup shows Scenario 1 (left) and Scenario 2 (right) with different access point locations, as well as the starting and ending (goal) points of the autonomous e-vehicles in the simulation trials.}}
    \label{fig:simulation}
    \vspace{-2mm}
\end{figure*}

\begin{figure}[htbp]
    \centering
    \includegraphics[width=\linewidth]{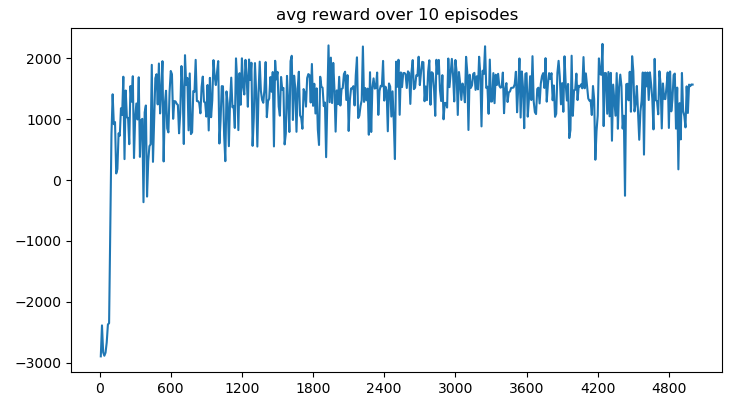}
        \vspace{-4mm}
    \caption{Convergence of the proposed approach based on training performed on simulations and the parameters of Table II.}
    \label{fig:reward_graph}
\end{figure}
\begin{figure}[htbp]
    \centering
    \includegraphics[width=\linewidth]{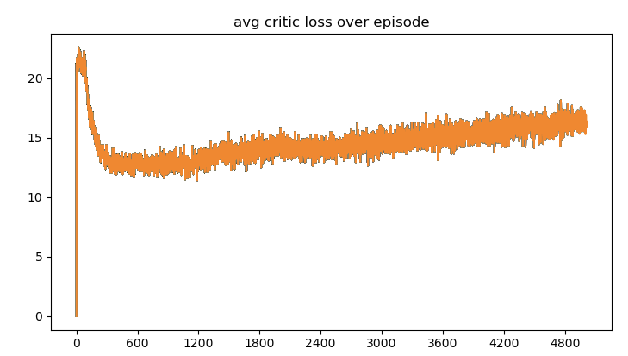}
        \vspace{-4mm}
    \caption{Performance of the critic loss function during training performed on simulations and the parameters of Table II.}
    \label{fig:critic_graph}
\end{figure}
\begin{figure}
        \centering
        \includegraphics[width=\linewidth]{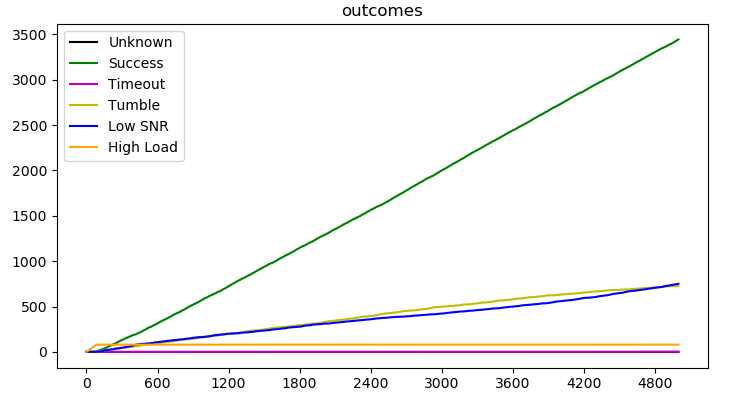}
        \vspace{-4mm}
        \caption{Outcomes recorded after each episode during the training phase.}
        \label{fig:outcomes}
\end{figure}

\section{Simulation Results}
We now present the simulation results to demonstrate the performance of the proposed DDPG-based AP selection.

\subsection{Simulation Settings}
To validate the proposed algorithm, we conducted our simulation experiments in the Gazebo simulator. The learning module was developed using the ROS2 framework, which leverages services and clients to establish a communication network between all the modules. Vehicles are represented as robots in the simulation space, with a specified number of vehicles and access points (APs). The vehicles are free to move within a 20x20 meter area, and in each trial, they are given a target point to navigate towards. Each vehicle maintains a linear velocity of 0.9 m/s. 

Throughout the simulation, the vehicle density remains constant. The APs are also mobile, but their speed is relatively slow to capture a highly dynamic scenario in the simulation setting. The minimum SNR threshold for all the AP associations is set to be 22 dBm, which represents the \textit{low SNR} condition, and the threshold for maximum AP-load is set to be equal to the total number of vehicles in the environment, which represents the \textit{High Load} condition. 

In the proposed DDPG-based AP-selection model, we design the actor-network with 3 fully connected layers with hidden layers of 512, each with ReLU and sigmoid activations. The critic network has 4 fully connected layers with ReLU activation.  We assume that there are 100 steps in each episode and the environment is reset at the beginning of each episode. We update the actor and the critic network with a learning rate of  0.003 each. The detailed training parameters are summarized in Table ~\ref{tab:training_parameters}.

\begin{table}[htbp]
    \centering
    \caption{DDPG TRAINING PARAMETERS}
    \label{tab:training_parameters}
      \vspace{-2mm}
    \begin{tabular}{lc}
        \toprule
        Parameter & Value \\
        \midrule
        Learning rate of actor and critic network & 0.003 \\
        Buffer size $\mathcal{B}$ & 1000000 \\
        Batch size $\mathcal{N}$ & 1024 \\
        Discount factor $\gamma$ & 0.99 \\
        Soft Update parameter $\tau$ & 0.003 \\
        Epsilon decay & 0.9999 \\
        Epsilon minimum & 0.05 \\
        Noise standard deviation $\sigma_{min}$, $\sigma_{max}$ & 0.1 \\
        \bottomrule
    \end{tabular}
\end{table}

We tested our approach and the baselines in two distinct scenarios (see Fig.~\ref{fig:simulation}), each differing in AP placements. 

\textbf{In Scenario 1}, the APs are placed close to each other in a single column. As vehicles traverse their coverage areas, they traditionally perform more handovers. This scenario tests our methodology's handover rate, thereby assessing our claim of maintaining consistent edge connections.

\textbf{In Scenario 2}, the APs are arranged in a square formation, occupying each corner. This scenario is designed to evaluate whether our proposed approach can maintain a stable connection throughout the trails. 

The proposed method is tested against the three legacy methods for AP associations in WiFi WLANs:
\begin{itemize}
    \item \textbf{Random Allocation (RA)}: In the RA method, vehicles randomly choose an AP when making handovers. This method does not take into account any system or environmental metrics.
    \item \textbf{Strongest Signal First (SSF)}: In the method, a vehicle associates with the AP from which it receives the maximum signal strength. RSSI (Received Signal Strength indicator) or signal-to-noise ration (SNR) could be taken into account for AP selection. A shortcoming of the SSF approach is that if an AP is overloaded, associating more stations with it can cause congestion, leading to increased packet loss and end-to-end delay.
    \item \textbf{Least Loaded First (LLF)}: In the LLF scheme, the least loaded AP is selected first. Selecting the least loaded AP provides load balancing across multiple APs; however, it may force a vehicle to associate with a distant AP, resulting in poor connection quality. 
\end{itemize}

The simulation results are discussed below, as well as the 100 trials conducted on each scheme.

\subsection{Convergence performance}
Fig ~\ref{fig:reward_graph}. demonstrates the learning performance of cumulative reward for all the vehicles over the course of 5000 episodes in the case of six vehicles and four APs. It can be observed that the value of the loss function is high at the beginning of the learning process. When the training is around 300 episodes, the loss value (Fig. ~\ref{fig:critic_graph}) declines to a very small and relatively stable value. This means that our approach achieves a good performance with fast convergence speed and a stable learning process. The Fig.~\ref{fig:outcomes}. represents the outcomes recorded after each episode during the learning process and the success rate of vehicles with optimum AP selection to arrive at the target locations.


\subsection{Average SNR}
\begin{figure}[t]
    \centering
    \includegraphics[width=1\linewidth]{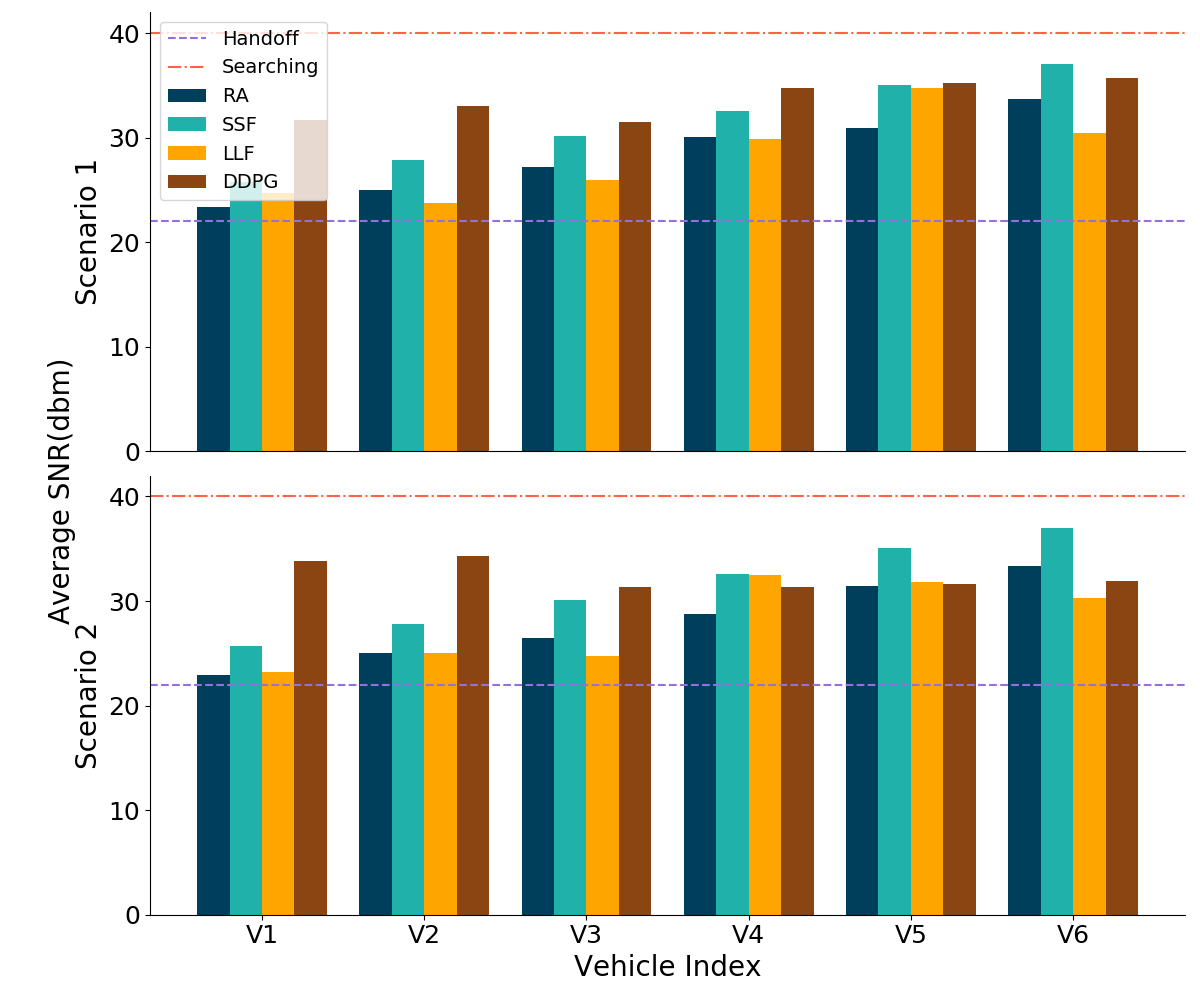} 
        \vspace{-4mm}
    \caption{Average SNR of vehicles between associated APs.}
    \label{fig:snr-graph}
\end{figure}

Fig.~\ref{fig:snr-graph} represents the average SNR of the links maintained by each vehicle. As observed in Scenario 1, our proposed DDPG-based solution maintains the highest average SNR, as expected. The SSF is a close second, as it follows a greedy approach that only connects the APs with the best signal quality. This approach does lead to establishing good connections with the APs but results in significant performance degradation due to overloading edge servers, where the congestion of fast data transfer leads to low computing performance. Both SSF and DDPG ensure stable communication links for all vehicles at all times, compared to the LLF and RA approaches, which fall behind in maintaining a good average SNR. In Scenario 2, DDPG maintains a good SNR for all vehicles, while SSF maintains an uneven SNR, with some vehicles being connected to very low-quality signals and others to high-quality signals. The RA method follows a similar pattern to SSF but with a lower SNR.

\subsection{Average AP-load}
\begin{figure}[t]
    \centering
    \includegraphics[width=\linewidth]{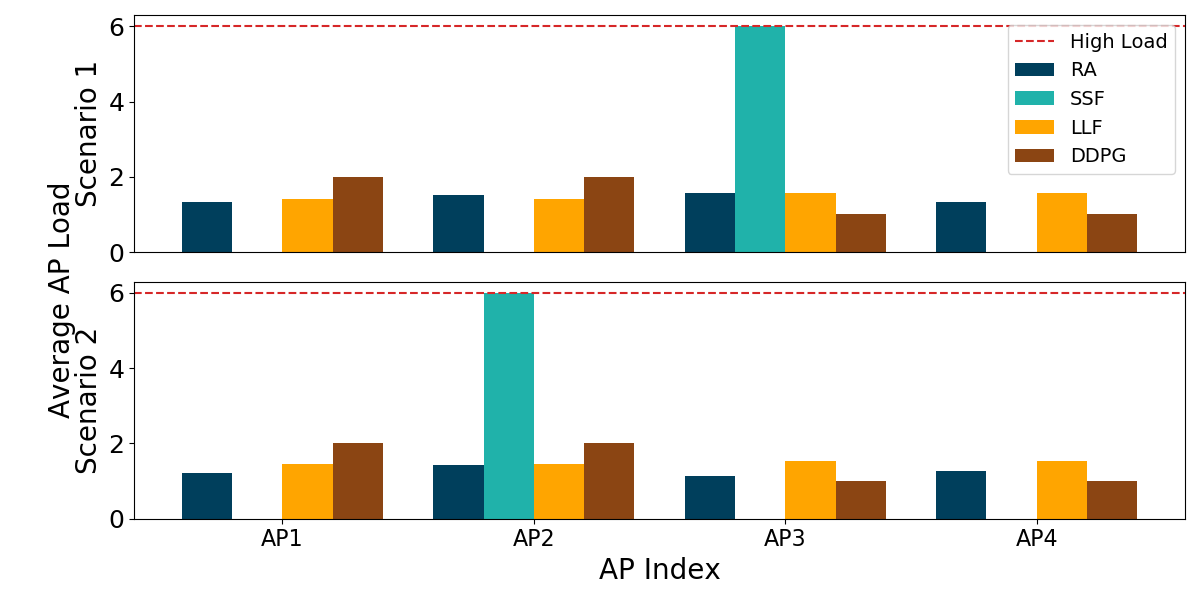} 
        \vspace{-6mm}
    \caption{Average AP-Load (number of APs = 4).}
    \label{fig:ap-load}
\end{figure}

Fig.~\ref{fig:ap-load} illustrates the average AP-load across all APs during the trials. It is evident that the DDPG algorithm maintains an optimal load distribution on the most frequently selected APs for both scenarios. In contrast, the SSF method concentrates the maximum load on the single AP with the highest signal strength value. Conversely, the LLF scheme appears to manage load balancing effectively by considering APs with the least load, thereby avoiding both over-utilization and underutilization. However, this approach may compel a vehicle to connect to a lower-quality AP, potentially resulting in dropped packets and data loss.
 \begin{figure}[t]
    \centering
    \includegraphics[width=1\linewidth]{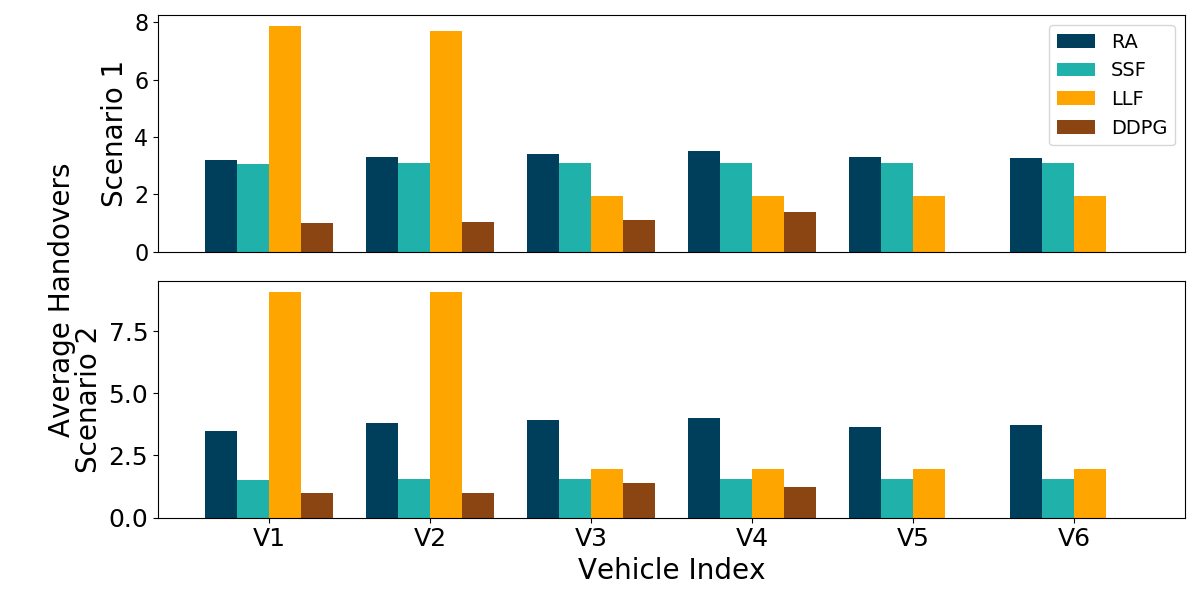} 
        \vspace{-6mm}
    \caption{Average Handover of vehicles between APs.}
    \label{fig:handover-graph}
\end{figure}
 \begin{figure}[t]
    \centering
    \includegraphics[width=1\linewidth]{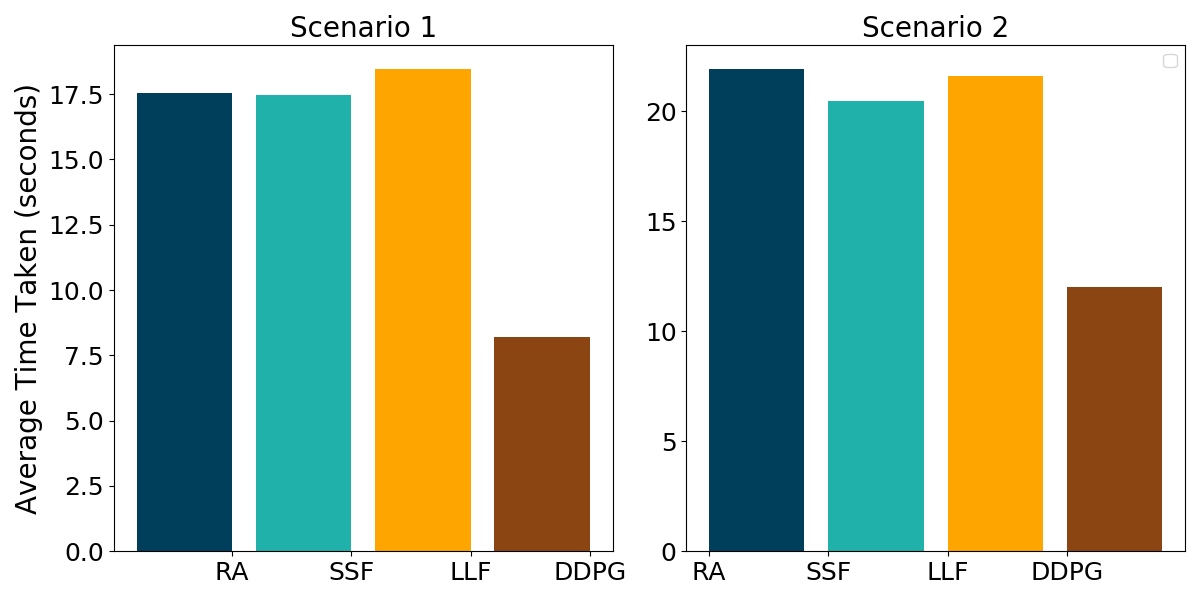} 
        \vspace{-6mm}
    \caption{Average time taken to complete navigation task.}
    \label{fig:time-graph}
\end{figure}

\subsection{Average handovers}
The proposed DDPG algorithm demonstrates superior handover performance compared to all baselines, as shown in Fig.~\ref{fig:handover-graph} for both scenarios. By minimizing the number of handovers over 100 trials, it maintains a well-balanced approach. On average, our methodology performs either no handovers or just one, ensuring consistent edge connections and reducing the need for task migrations while vehicles navigate longer distances, all while maintaining good link quality. In comparison, the SSF method averages three handovers in Scenario 1 and between zero and two in Scenario 2. The LLF scheme, however, performs frequent handovers, resulting in high transfer and service latency.

\subsection{Average time taken}
Additionally, the average time, as shown in Fig.~\ref{fig:time-graph}, required to complete the trials, indicates that random allocation takes the longest to complete. Both SSF and LLF exhibit similar performance in terms of trial duration, but the DDPG algorithm shows a shorter timeframe. This efficiency is attributed to DDPG's ability to select the most optimal APs based on network metrics, resulting in fewer handovers and smoother transitions between APs. The results indicate that the selection of the optimal AP based on network and load dynamics reduces the latency of time-sensitive tasks, thus reducing the transfer latency associated with the edges.

\section{Conclusion}
This study addresses a notable gap in the existing literature on minimizing handovers and optimizing resource allocation in vehicular edge networks. We present a DDPG-based learning framework for continuous AP-selection and a joint optimization approach that enhances network stability and load management while meeting QoS requirements. Simulation results demonstrate the superior performance of our proposed policy compared to baseline algorithms, highlighting its potential to significantly improve network efficiency and service quality in vehicular environments.

\bibliography{references}
\bibliographystyle{IEEEtran}
\end{document}